# The Rise of Internet of Things (IoT) in Big Healthcare Data: Review and Open research Issues

Zainab Alansari[1,2], Safeeullah Soomro[1], Mohammad Riyaz Belgaum[1] and Shahaboddin Shamshirband[2]

[1] College of Computer Studies, AMA International University,
Kingdom of Bahrain
{zeinab, s.soomro, bmdriyaz}@amaiu.edu.bh

[2] College of Computer Science and Information Technology,
University of Malaya, Malaysia
shamshirband@um.edu.my

**Abstract.** Health is one of the sustainable development areas in all of the countries. Internet of Things has a variety of use in this sector which was not studied yet. The aim of this research is to prioritize IoT usage in the healthcare sector to achieve sustainable development. The study is an applied descriptive research according to data collection. As per the research methodology which is FAHP, it is a single cross-sectional survey research. After data collection, the agreed paired comparison matrices, allocated to weighted criteria and the priority of IoT usage were determined. Based on the research findings, the two criteria of "Economic Prosperity" and "Quality of Life" achieved the highest priority for IoT sustainable development in the healthcare sector. Moreover, the top priorities for IoT in the area of health, according to the usage, were identified as "Ultraviolet Radiation", "Dental Health" and "Fall Detection".

**Keywords:** Internet of Things (IoT); Healthcare; Fuzzy-AHP; Big Data

## 1 Introduction

Internet revolution in the recent decades showed that the new kinds of technologies can affect all aspects of the businesses [1]. Nowadays, with the help of new technologies such as network connections, wireless communication, and sensors, ubiquitous communication is always possible. Moreover, Business owners reflected many articles regarding IoT and introduced it as new solutions in ICT which they believe it has the potential to earn a great and valuable income.

The purpose of IoT is empowering the objects for connection at anytime, anywhere, with anything and any person that get the ideal use of any route, service or network. Internet of Things is a new evolution of the Internet. IoT is a new technology that focuses on the environmental effect and deals with a variety of wireless and wired connections which are communicating with each other. These objects are working together to develop an application for a new service and achieve a common goal together. In fact, it considers as a development challenge for creating a great smart world. A world which in its actual shape is digital and virtual but moves toward the development of smart world which creates smarter areas of energy, transportation, health, cities, and many more [2] and [3].

In practice, different countries have numerous motivations to support IoT including United States [4], China [5], European Union [6] and India [7]. According to the report from IoT European Research Cluster (IERC), three motivations for the development of IoT in countries is Economic Prosperity, Quality of Life and Environmental Protection [8]. It has been discussed the sustainable development literature [9].

No priority of IoT has been determined in the health sector, and it seems to be essential to prioritize some areas of IoT which have the highest potential for sustainable development in the health sector. Furthermore, the use of innovative technologies has always been considered by researchers, but no research has been conducted on IoT in the health sector so far. The aim of this study is to prioritize the functional areas of the health sector which is the development of IoT to achieve sustainability in the health sector. Therefore, the study seeks to answer the following questions:
- How much is each indicator's weight of Economic Prosperity, Quality of Life and Environmental Protection, to assess the IoT in the health sector?
- What is the priority of each IoT application in the health sector?

## 2 Literature Review

The Internet of Things for the first time was used in 1999 which described the world where anything, including people, animals, plants, and even inanimate objects (such as cars), being able to have their digital identity which is allowing the computers to organize and manage them.

Nowadays the internet is connecting all the people, but the IoT is connecting all the objects together, and the people can control and manage them by the use of available applications in smartphones and tablets [10]. Actually, IoT is a new concept in the technology and communication world which considered as a modern technology provides the capability of sending data for anything (human, animal, or object) via network connection rather Internet or an intranet [11].

Businesses have focused heavily on IoT, and this has led to the Electronic Business (e-Business) development [12] and in many cases customer relationship management is easier through IoT [13]. In fact, IoT is an approach that will improve the interoperability between an object with object, object with human, and human with an object and with the help of such an approach the new services will appear [14]. Moreover, one of the primary objectives of IoT is to increase intelligence in life, business, and economy [15].

### 2.1 Smart Health

IoT promises market potential in the field of e-health services and the telecommunications industry [16]. IoT can enhance business intelligence in hospitals and ease serving the patients in health sectors [17]. The Internet of Things can improve the health grounds and prevent diseases by providing ongoing monitoring activities to ordinary people or prone patients [16] and [18]. Furthermore, it empowers the patients and helps the businesses to profit from this new innovative in the market. Somehow it improves the patient's Social problems, people's concerns about health and the quality of their lives [19] also contributes to economic prosperity in health sector [20]. With the help of this technology the hospital activities impact on the environment (Such as production and elimination of hospital waste) can be better managed and less likely to damage the environment [21].

The usages of IoT can develop some platforms which provide smart and innovative services to patients and people in need of medical attention. Furthermore, improves their health, Security, ease of access to emergency medical care, continuing care and quick support also improving the quality of life [16].

## 2.2 Applicable Areas of IoT in The Health Sector

IoT European Research Cluster (IERC), has been presented a comprehensive classification of relevant areas of IoT in smart health in the health sector. Some usages are of the type of services, and some are a kind of product. Related Areas of IoT in The Health Sector (smart health) are:
• Fall Detection: This usage focused on helping the physically challenged and elderly people in their lives so that they can live independently;
• Medical Fridges (Internal temperature protective control): Some organic elements must be kept in containers with certain conditions (temperature). IoT can well assume this task and Cause objects interaction;
• Sportsmen Care: The application used to measure the Weight, sleep, exercise, blood pressure and other relevant parameters for professional athletes;
• Patient Surveillance: used for remote in-hospital monitoring, (especially the elderly) or used for patient's home care;
• Chronic Disease Management: Taking care of patients with chronic diseases while there is no need of physical attendance. This technology reduces the presence of people in hospitals and results in lower costs, reduces hospital stay and reduces traffic (Even reduces fuel consumption);
• Ultraviolet Radiation: UV rays Measurement and notifying the people not to enter certain areas or refrain of exposure to UV rays at certain hours;
• Hygienic Hand Control: By linking devices such as designed RFID for emissions measurement, environmental pollution could be identified;
• Sleep Control: Devices that by linking to individuals, identifies some signs such as heart rate, blood pressure during sleep and the data may be collected and will be analyzed after;
• Dental Health: Bluetooth-enabled Toothbrush with the help of smartphone apps records someone's brushing information to study the person's brushing habits and share the statistics with the dentist [16].

### 2.3 IoT Sustainability Indicators in Health Sector

United Nations [22] described the sustainability concept that seeks to define human needs without compromising the ability of future generations to meet their needs. On the other hand, Porter and Kramer [23] stated that businesses are the primary cause of the social, environment and economy problems in recent years. Lack of business confidence causes the political leaders to develop policies that weaken competitiveness and economic growth. Selection of such policies over the last few decades created the impression that the "Economic Prosperity" is a reverse of "Social progress".

The government must learn the management in such way to ease the creation of shared social and economic value, rather than to stop it. Creating simultaneously shared value focuses on the relationship between social and economic progress and has the power of encouraging the next wave of global growth [24]. Porter and Kramer [23] stated that the idea of creating shared value, clarifies the government's role in responsible behavior towards the sustainable development.

In this study, Economic Prosperity, Quality of Life and Environmental Protection criteria has been used to assess the IoT scope in the health sector. As previously mentioned, IERC introduced this criterion as drivers for the IoT development [25], and it is considered in sustainable development literature as well [26]. In the research literature, criteria of economic prosperity were placed in the financial domain [27] Quality of life were set in the social sphere [28], and environmental protection included in an environmental field [29].

## 3 Conceptual Model

Figure 1 shows the Conceptual Model of the study which prioritized IoT in health sector according to sustainable development criteria and based on the Fuzzy Analytical Hierarchy Process (FAHP) method:

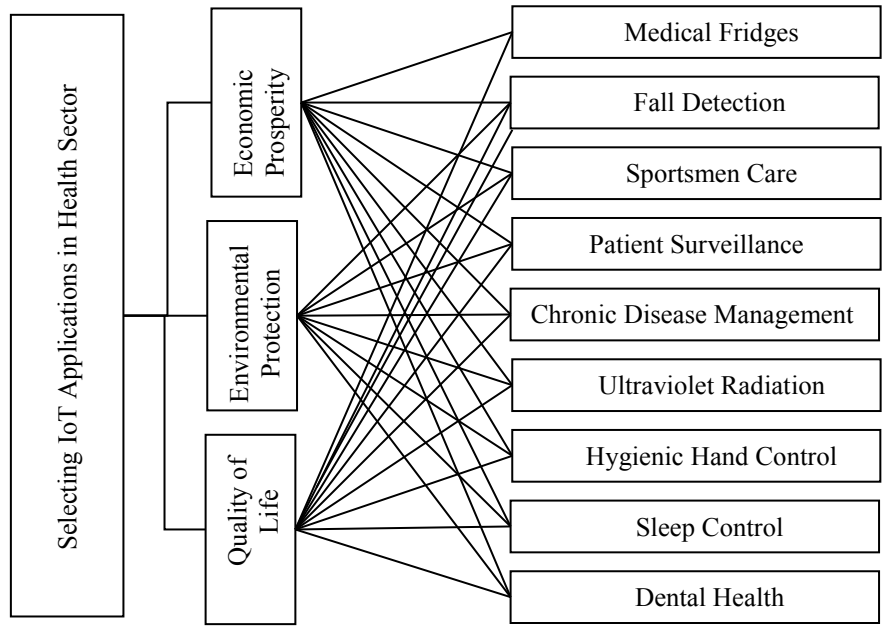

**Fig. 1.** The Conceptual Model of the study based on FAHP.

## 4 Methodology

The study objective is an applied research while in terms of data gathering it is considered as a descriptive non-experimental among the quantitative researches. Since Fuzzy Analytic Hierarchy Process (FAHP) is used for weighting and prioritization of IoT in the health sector (Smart Health), it is a Cross-sectional study among the descriptive studies.

Initially, each of the Economic Prosperity, Quality of Life and Environmental Protection criteria were weighted. Then in three Pair-Wise Comparison Questionnaire, each application of IoT in the health sector was compared separately according to their criteria. Finally, the Decision Matrix was obtained. Content validity method used to assess the validity of questionnaire and twelve experts in IoT Health Sector confirmed the decision-making matrix components such as criteria and options. The statistical population of the study consisted of experts who are familiar with IoT that has a background of trade cooperation or was business partners to provide services or advice to hospitals and health centers in the use of new technologies.

Due to the limited number of experts, the snowball method was used; Therefore, after referring to the Communication Research Centre, Experts in the health field were identified as a pioneer of IoT research objects. A joint meeting was handled with IoT experts to reach an agreement on each of the paired comparisons. Twenty experts have invited which only twelve experts attended the meeting and the obtained Pair-Wise Comparison Questionnaire were the base on data analysis of this study.

## 4.1 Fuzzy Analytical Hierarchy Process (FAHP)

AHP model is a Multi Attribute Decision Making (MADM) that professor saaty proposed in 1970 [30]. In 1996, a Chinese researcher named Chang proposed a fuzzy AHP method based on development analysis [31]. In fact, FAHP method was developed based on AHP and fuzzy logic [32]. Triangular Fuzzy Numbers are the numbers used in this method. The geometric space in such a fuzzy environment is shown in Figure 2:

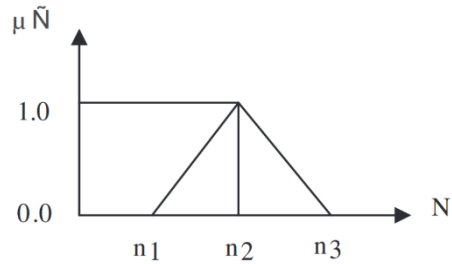

**Fig. 2.** A Triangular Fuzzy Number, N [32].

In this method, the membership function and fuzzy scale defined in Table 1:

**Table 1.** Membership functions and the definition of fuzzy scale [32].

| Intensity of importance | Fuzzy number | Definition | Membership function |
|---|---|---|---|
| 9 | $\tilde{9}$ | Extremely more importance (EMI) | (8, 9, 10) |
| 7 | $\tilde{7}$ | Very strong importance (VSI) | (6, 7, 8) |
| 5 | $\tilde{5}$ | Strong importance (SI) | (4, 5, 6) |
| 3 | $\tilde{3}$ | Moderate importance (MI) | (2, 3, 4) |
| 1 | $\tilde{1}$ | Equal importance (EI) | (1, 1, 2) |

The stages of applying FAHP method in this research are as follows:
1. Evaluation of literature research to determine the sustainability criteria and the use of IoT applications in health sector (smart health);
2. Formation of a decision team to examine the questionnaire validity;
3. Distributing the questionnaires and creating paired comparisons matrix with Fuzzy terminology for each questionnaire;
4. Weighting of criteria and determining the scores of the options for each of the criteria using FAHP;
5. Ranking the usage of IoT in the health sector by FAHP method.

The calculation of FAHP method to achieve the weight of criteria and prioritization of IoT applications in the health sector conducted with Chang [31] method.

## 5 Analysis and Results

Determining the weights of criterion and the score of each were completed after conducting several meetings to fill the agreed questionnaire. Four paired comparison questionnaire (one questionnaire to compare paired criterion and three questionnaires to compare the IoT applications in health sector according to each criterion) with the help of fuzzy words (Table 1). Using the FAHP, criterion's weight obtained from the paired comparison agreed on matrices as shown in Figure 3.

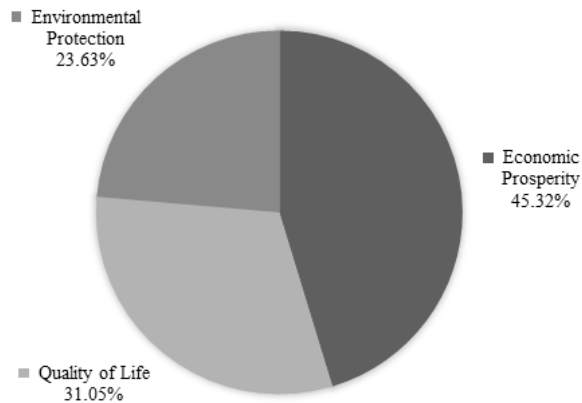

**Fig. 3.** Weight of Criteria Sustainability.

Similarly, each option in each of the criteria calculated using FAHP, and decision matrix was formed as shown in Table 2.

**Table 2.** Decision matrix derived from the average weight of each criterion for each option.

| Options | Economic Prosperity | Quality of Life | Environmental Protection |
|---|---|---|---|
| Fall Detection | 0.109 | 0.096 | 0.14 |
| Medical Fridges | 0.084 | 0.11 | 0.039 |
| Sportsmen Care | 0.069 | 0.0836 | 0.153 |
| Patient Surveillance | 0.117 | 0.0954 | 0.121 |
| Chronic Disease Management | 0.079 | 0.104 | 0.025 |
| Ultraviolet Radiation | 0.193 | 0.132 | 0.176 |
| Hygienic Hand Control | 0.098 | 0.094 | 0.12 |
| Sleep Control | 0.068 | 0.143 | 0.059 |
| Dental Health | 0.183 | 0.142 | 0.167 |

### 5.1 Ranking the options using FAHP

Ultimately, with the help of weights obtained for sustainable development criteria (Figure 3) and decision matrix (Table 2), The final score and rank of each IoT application in health sector achieved using the FAHP as shown in Table 3.

**Table 3.** Scores and the Priority of IoT in the Health Sector.

| Options | Rank | Score |
|---|---|---|
| Ultraviolet Radiation | 1 | 0.0567 |
| Dental Health | 2 | 0.0555 |
| Fall Detection | 3 | 0.0374 |
| Patient Surveillance | 4 | 0.0371 |
| Hygienic Hand Control | 5 | 0.0340 |
| Sportsmen Care | 6 | 0.0311 |
| Sleep Control | 7 | 0.0297 |
| Medical Fridges | 8 | 0.0271 |
| Chronic Disease Management | 9 | 0.0247 |

## 5 Conclusions

According to the results, the most important criteria of IoT in the health sector for sustainable development is economic prosperity with a weight of 45.32% and then the quality of life with the weight of 31.05%. According to the Experts, the weight of Environmental Protection to develop IoT smart health sector is 23.63%. Therefore, it is recommended to the policy makers of the health sector to focus on developing the new technologies such as IoT application on economic criteria like employment and revenue, then social criteria such as increasing the welfare of patients and citizens and satisfaction of hospital personnel in the use of medical tools. Moreover, the environmental impact of these technologies in the environmental sector such as probable radiation, harmful radio waves, preventing the creation of waste and water should not be forgotten.

Based on Table 2, If only the criteria of Economic Prosperity are concerned in IoT field, the "Ultraviolet Radiation", "Dental Health" and "Patient Surveillance" are in priority. According to the same results, if only the criteria of Quality of Life is considered, "Sleep Control", "Dental Health" and "Ultraviolet Radiation" are in top priority and improve the citizen's quality of life more than another criterion. Finally, if only the criteria of Environmental Protection are considered, "Ultraviolet Radiation", "Dental Health" and "Sportsmen Care" are in top priorities more than any other are of IoT in the health sector or smart health contribute to the improvement of environmental protection.

However, in this study, all the three criterion are effective based on their importance in the IoT health sector. The ranking results based on FAHP of the expert's opinion and the basis for paired comparisons of IoT applications in health sector and according to the triple sustainability criteria which were obtained in this research (Table 3), indicate that the priority of IoT in health sector using sustainable development criteria are "Ultraviolet Radiation", "Dental Health", "Fall Detection", "Patient Surveillance", "Hygienic Hand Control" and "Sportsmen Care". "Sleep Control", "Medical Fridges" and "Chronic Disease Management" are located in the last rank. Therefore, it is recommended to the government and relevant health centers to give more support to the IoT application in "Ultraviolet Radiation", "Dental Health" and "Fall Detection" as they followed the most interests of stability.

## 6   Future Studies

The results of this study increase the knowledge of IoT, familiarity of IoT innovation in the health sector, and encourages the usage of new technologies according to the sustainable development criterion. However, this study has some limitations, since the IoT applications experience in the health sector is limited, it may influence the motivation of investment based on given prioritization which seems the need for technical and economic feasibility studies. Moreover, the governments are still not supporting the IoT governance, its regulation and consumer and producer's rights. Therefore, it could be a future studies concern of IoT implementation.

## References

1. Premkumar, G. & Roberts, M.: Adoption of new information technologies in rural small businesses. In: Omega 27, no. 4, pp. 467--484. (1999)
2. Atzori L, Iera A, Morabito G.: The internet of things: A survey. In: Computer networks, pp. 2787--805. (2010)
3. Sundmaeker, H., Guillemin, P., Friess, P. & Woelfflé, S.: Vision and challenges for realising the Internet of Things. In: CERP- IoT – Cluster of European Research Projects on the Internet of Things. (2010)
4. U.S. Government Promoting Development Of The "Internet Of Things", http://smartamerica.org/news/u-s-government-promoting-development-of-the-internet-of-things
5. China Internet of Things, http://iot.cqna.gov.cn
6. IERC (European Research Cluster on the Internet of Things), http://www.internet-of-things-research.eu/about_iot.htm
7. Deity (Department of Electronics & Information Technology), http://deity.gov.in/content/draft-internet-thingsiot-policy
8. Smith, I.G., C.: The Internet of Things 2012 New Horizons. European Research Cluster on the Internet of Things, UK (2012)
9. Carter, C. R. & Easton, P. L.: Sustainable supply chain management: evolution and future directions. In: International Journal of Physical Distribution & Logistics Management, 41(1), pp. 46 -- 62. (2011)
10. Ashton, K.: That 'internet of things' thing. In: RFiD Journal, 22(7), pp. 97--114. (2009)
11. Chui, M., Löffler, M. & Roberts, R.: The internet of things. In: McKinsey Quarterly, 2, pp. 1--9. (2010)
12. Uckelmann, D., Harrison, M. & Michahelles, F.: Architecting the internet of things. In: Springer Science & Business Media. (2011)

13. Xiaocong, Q. & Jidong, Z.: Study on the structure of "Internet of Things (IOT)" business operation support platform. In: Communication Technology (ICCT), 12th IEEE International Conference, pp. 1068--1071. (2010)
14. Miorandi, D., Sicari, S., DePellegrini, F.and Chlamtac, I.: Internet of things: Vision, applications and research challenges. In: Ad Hoc Networks, 10(7), pp. 1497--1516. (2012)
15. Tan, L., & Wang, N.: Future internet: The internet of things. In: Advanced Computer Theory and Engineering (ICACTE), 3rd International Conference on. 5: V5-376. IEEE. (2010)
16. Vermesan, O. & Friess, P.: Internet of Things-From Research and Innovation to Market Deployment. In: River Publishers. (2014)
17. Roman, R., Najera, P. & Lopez, J.: Securing the internet of things. In: Computer, 44(9), pp. 51--58. (2011)
18. Bandyopadhyay, D. & Sen, J.: Internet of things: Applications and challenges in technology and standardization. In: Wireless Personal Communications, 58(1), pp. 49--69. (2011)
19. Helal, A., Cook, D. J. & Schmalz, M.: Smart home-based health platform for behavioral monitoring and alteration of diabetes patients. In: Journal of diabetes science and technology, 3(1), pp. 141--148. (2009)
20. Haller, S., Karnouskos, S. & Schroth, C.: The internet of things in an enterprise context. In: Springer Berlin Heidelberg. pp. 14—28. (2009)
21. Perera, C., Zaslavsky, A., Christen, P. & Georgakopoulos, D.: Context aware computing for the internet of things: A survey. In: Communications Surveys & Tutorials, IEEE 16.1, pp: 414 -- 454. (2014:a)
22. United Nations (UN); UN Documents: Gathering a body of global agreements, http://www.un-documents.net/wced-ocf.htm
23. Porter, M. E. & Kramer, M. R.: Creating shared value. In: Harvard business review, 89 (1/2), pp. 62--77. (2011)
24. Crane, A., Palazzo, G., Spence, L. J. & Matten, D.: Contesting the value of "creating shared value". In: California management review, 56(2), pp. 130--153. (2014)
25. Smith, I.G.: The Internet of Things 2012 New Horizons. In: European Research Cluster on the Internet of Things, Halifax, UK. (2012)
26. Carter, C. R. & Easton, P. L.: Sustainable supply chain management: evolution and future directions. In: International Journal of Physical Distribution & Logistics Management, 41(1), pp. 46--62. (2011)
27. Bansal, P.: Evolving sustainably: a longitudinal study of corporate sustainable development. In: Strategic management journal, 26(3), pp. 197--218. (2005)
28. Baud, I. S. A., Grafakos, S., Hordijk, M. & Post, J.: Quality of life and alliances in solid waste management. In: contributions to urban sustainable development. Cities, 18(1), pp. 3-12. (2001)
29. Zhang, K. M and Wen, Z. G.: Review and challenges of policies of environmental protection and sustainable development in China. In: Journal of environmental management, 88(4), pp. 1249--1261. (2008)
30. Saaty, T. L.: What is the analytic hierarchy process? In: Springer Berlin Heidelberg, pp. 109--121. (1988)
31. Chang, D. Y.: Applications of the extent analysis method on fuzzy AHP. In: European journal of operational research, 95(3), pp. 649--655. (1996)
32. Büyüközkan, G., Çifçi, G. & Güleryüz, S.: Strategic analysis of healthcare service quality using fuzzy AHP methodology. In: Expert Systems with Applications, 38(8), pp. 9407--9424. (2011)